# Subwavelength and microstructured planar THz waveguides, theory and experiment

A. Markov[1], A. Mazhorova[1], B. Ung[1], H. Minamide[2], Y. Wang[2], H. Ito[2], and M. Skorobogatiy[1]

1. École Polytechnique de Montréal, Génie Physique, Québec, Canada
2. RIKEN, Sendai, Japan.
[*]*maksim.skorobogatiy@polymtl.ca*

**Abstract:** Planar dielectric subwavelength waveguides are proposed as waveguides for THz radiation. The waveguide porous design maximizes the fraction of power guided in the air to avoid the complexity that all the materials are highly absorbent in the THz region. Convenient access to the mode region makes them also useful for sensing of biological and chemical specimens in the THz region.

## References and links

## 1. Introduction

The main complexity in designing terahertz waveguides is the fact that almost all materials are highly absorbent in the terahertz region [1]. Since the lowest absorption loss occurs in dry air, an efficient waveguide design must maximize the fraction of power guided in the air. The subwavelength type of waveguides has been proposed based on this concept. The constituent dimensions of such a waveguide are much smaller than the wavelength and, thus, it has high power fraction outside the lossy material.

We propose a novel design of subwavelength low-loss in THz regime waveguide which takes a form of a planar composite waveguide. Its structure consists of multiple layers of thin (25-50 μm) polyethylene film which is highly absorbing in this spectral range and, hence, is not suitable for regular step index waveguides. These layers are stacked together but separated by air spacing of comparable thickness. Such planar waveguides have been proposed not only for THz wave delivery but also for sensing of biological and chemical specimens in the terahertz region.

There are three main advantages of this kind of waveguides: first, the major portion of THz power launched into such a waveguide is confined within the air layers, secondly, modal refractive index of porous waveguide is smaller compared to pure polymer, thirdly, and porous waveguides exhibit smaller transmission losses than bulk material.

## 2. Waveguide design

The photograph of the planar composite waveguide is presented in the Fig. 1 (a). It has been fabricated from multiple layers of commercially available 50 μm thickness polyethylene film. The layers were stacked together and separated by layers of adhesive tape by the edges. The edges were clutched tightly by metal holder, thus, making air gaps between the individual polyethylene layers. The thickness of the adhesive tape has been measured with the optical microscope and has been found to be approximately 75 μm (see Fig. 1 (b)), resulting in the same thickness of the air separating layers of the waveguide. The produced waveguide width is 3.3 cm and its length is around 11 cm. The maximal investigated number of polyethylene layers is 10, resulting in 1.2 mm overall thickness of the waveguide. The influence of the metal borders of the waveguide on its optical properties will be discussed in the appropriate part of the paper.

Polyethylene film is highly absorbing in THz spectral range, therefore, the major portion of THz power launched into such a waveguide is confined within the air layers. The waveguide optical properties are sensible to the permittivity of the separating layers, thus, making it possible to use the waveguide for sensing the specimen placed inside the waveguide structure. Modal refractive index of porous waveguide is smaller compared to pure polymer and it can be controlled by the choice of the thickness of the separating layers, thus, making the waveguide suitable for coupling applications. Also, the waveguide takes the advantages of its porosity in exhibiting smaller transmission losses than bulk material.

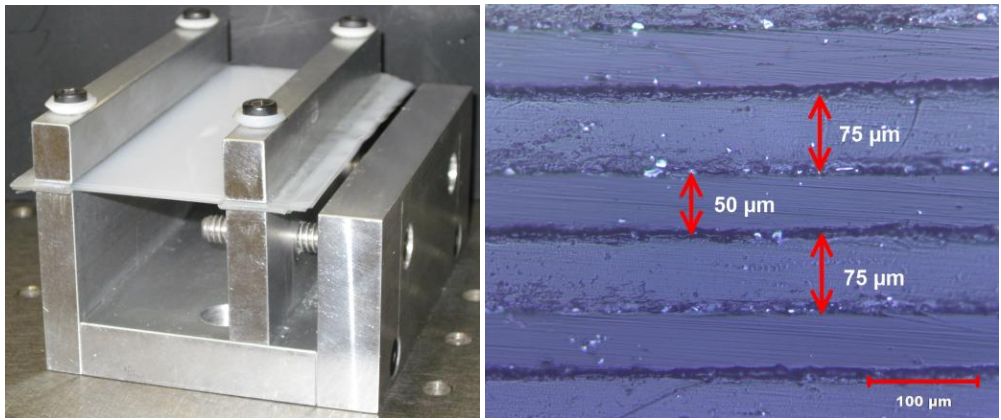

Fig. 1. (a) Photograph of the planar composite waveguide fabricated from polyethylene, (b) Microscope image of the waveguide, the thickness of the polyethylene layers is 50 μm, and separating layer thickness is 75 μm.

## 3. Bulk polymer material: refractive index and absorption losses

For modeling the optical properties of the waveguide, first, we have measured the transmission and absorption through the commercial polyethylene bulk material used in the waveguide fabrication. Characterization of refractive index and absorption losses was performed with a THz-time domain spectroscopy (TDS) setup using thick polymer slabs with parallel interfaces. The sample was prepared by cutting and polishing a 1.5 cm thick slice of the rod.

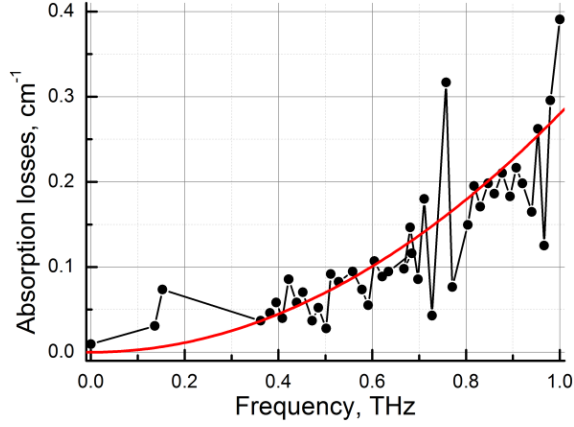

Fig. 2 Absorption losses of the bulk polyethylene. Red line – quadratic fitting curve. $\alpha_{PE}[cm^{-1}] \approx 0.28 \cdot f^2 [THz]$

The refractive index and absorption losses of polyethylene were retrieved by fitting the predictions of a transfer matrix model to the experimental transmission data [2]. The determined refractive index is largely constant between 0.10 and 1.00 THz and equal to $n_{PE}$ = 1.514. Power absorption losses in $cm^{-1}$ of polyethylene increase quadratically as a function of frequency and can be fitted as: $\alpha_{PE}[cm^{-1}] \approx 0.28 \cdot f^2 [THz]$ where f is the frequency in THz. The absorption losses rapidly increase with the increase of the frequency, the propagation length (defined as 1/e decay of intensity) in bulk plastic is about 3 meters for 0.1 THz frequency and only 3 cm for 1 THz.

## 4. Fundamental mode of the waveguide

The multilayer waveguides feature broader range of the single mode propagation regime compared to the slab waveguide of the same dimensions. The criterion for a single mode operation of the single slab waveguide:

$$k \cdot d \cdot \sqrt{n_2^2 - n_1^2} < \pi$$

where $n_2$ is the refractive index of the core material, which is typically larger than n = 1.45, and $n_1$ is the refractive index of the cladding material, usually equal to n = 1. Thus, for single mode propagation the criterion is $\lambda > 2 \cdot d$. On the other hand, the beam waste size depends on the wavelength of radiation. The theory predicts and the conducted experiments confirm that the beam diameter is directly proportional to the wavelength, and thus inversely proportional to the frequency. Particularly, in our THz setup we have measured that beam FWHM is about $2.5 \cdot \lambda$. For high coupling efficiency the waveguide width should exceed the beam waist, and hence, $d > 2.5 \cdot \lambda$, which contradicts the above mentioned criterion for single mode propagation. Thus, in a single slab waveguide one has to make a choice between the coupling efficiency and single mode propagation.

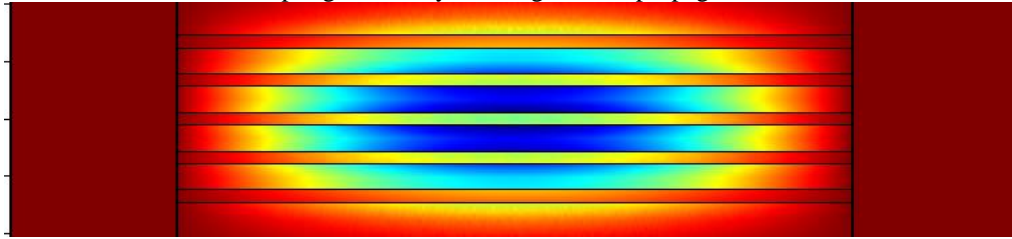

Fig. 3 Transverse electric field distribution of the fundamental mode of the waveguide

As for multilayer porous waveguide, it can still be low loss and single-mode; this can be accomplished by choosing a small enough size of the high-refractive-index layer. For example, a 1 mm wide waveguide composed of 10 layers of the commercially available polyethylene film with 25 μm thickness separated by 75 μm air gaps operates in single mode regime in wavelength range λ>700 μm, while 1 mm wide single slab waveguide of the same material is single-mode only for λ>2000 μm.

## 5. Details of the numerical modeling of the fields

The distribution of the transverse E-field components $\vec{E}_{output} = (E^x_{output}, E^y_{output})$ at the output facet of a waveguide of length $L_w$ is modeled as the coherent superposition of N guided modes:

$$\vec{E}_{output}(x, y, \omega) = \sum_{m=1}^{N} C_m \cdot \vec{E}_m(x, y, \omega) e^{i\frac{\omega}{c} \cdot n_{eff,m} \cdot L_w} \cdot e^{-\frac{\alpha_m L_m}{2}}$$

where $\vec{E}_{output} = (E^x_{output}, E^y_{output})$ stands for the transverse field components of the m-th guided mode. The variables $\alpha_m$ and $n_{eff,m}$ denote respectively the power loss coefficient and the real effective index of the m-th mode at a given frequency ω. The variable $C_m$ refers to the normalized amplitude coupling coefficients computed from the overlap integral of the respective flux distributions of the m-th mode with that of the input beam. Specifically, the definition of $C_m$ is based on the continuity of the transverse field components across the input interface (i.e. cross-section of the subwavelength waveguide) between the incident beam and the excited fiber modes:

$$C_m = \frac{\frac{1}{4} \int_{wc} dxdy \left( \mathbf{E}^*_m(x,y) \times \mathbf{H}_{Input}(x,y) + \mathbf{E}_{Input}(x,y) \times \mathbf{H}^*_m(x,y) \right)}{\sqrt{\frac{1}{2} \text{Re} \int_{wc} dxdy \left( \mathbf{E}^*_m(x,y) \times \mathbf{H}_m(x,y) \right)} \sqrt{\frac{1}{2} \text{Re} \int_{wc} dxdy \left( \mathbf{E}^*_{Input}(x,y) \times \mathbf{H}_{Input}(x,y) \right)}}$$

To model the input source, we assume a y-polarized beam with Gaussian shape along y-direction and uniform in x-direction whose fields are fields normalized to carry power P:

$$\vec{E}_{Input}(x, y) = \vec{x} \cdot \sqrt{\frac{2P}{\pi \sigma^2}} \cdot \exp\left[-\frac{y^2}{2\sigma^2}\right]$$

$$\vec{H}_{Input}(x, y) = \vec{y} \cdot \sqrt{\frac{2P}{\pi \sigma^2}} \cdot \exp\left[-\frac{y^2}{2\sigma^2}\right]$$

where the Gaussian beam waist parameter σ is related to the full-width half-maxima through $FWHM = 2\sigma\sqrt{2 \cdot \ln 2}$. The frequency dependence of the beam waist was measured independently and then fitted by a linear function of the input wavelength $\sigma \approx 2.5 \cdot \lambda$. This model was subsequently used in the following simulations.

## 5. Measurement setup

All the data in our experiments was acquired using a modified THz-TDS (Time-Domain Spectroscopy) setup. The setup consists of a frequency-doubled femtosecond fiber laser (MenloSystems C-fiber laser) used as a pump source and identical GaAs dipole antennae used as source and detector yielding a spectrum ranging from ~0.1 to 3 THz.

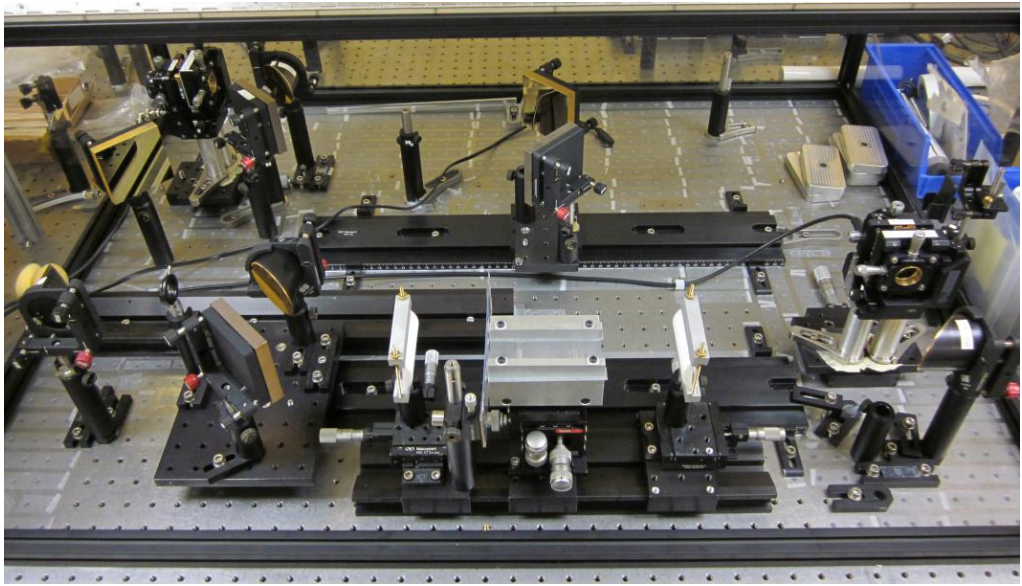

Fig. 4. Experimental modified THz-TDS setup with the waveguide fixed between two cylindrical lenses forming .

Contrary to the standard THz-TDS setup where the configuration of mirrors is static, our setup has mirrors mounted on translation rails. This flexible geometry facilitates mirrors placement, allowing measurement of waveguides up to 45 cm in length without realigning the setup. Fig. 4 illustrates the experimental setup where the planar waveguide is placed between the focal points of the two Teflon cylindrical lenses. The cylindrical lens changes the transverse profile of the beam at the entrance to the waveguide making the beam uniform along extensive horizontal side of the waveguide and remaining the beam Gaussian along the short vertical side of the waveguide.

## 6. Waveguide transmission and loss measurements

The transmission characteristics of the waveguide were measured using the THz-TDS imaging setup described in Section 5. The near-field probe was not scanned over the whole output facet of the waveguide. Instead, the probe remained positioned at a single spot located in the center (x0, y0) of the waveguide. We would like to note that this approach for measuring waveguide transmission loss is somewhat different from a traditional one that measures the total power coming out of the whole waveguide cross-section. Particularly, by using a near field probe placed in the waveguide center one preferentially measures losses of the few lowest order modes. This is related to the fact that such modes have their intensity maxima in the vicinity of the fiber core, thus providing the dominant contribution to the total transmission. We also note that the recorded transmission spectrum is sensitive to the exact (x0, y0) location of the near-field probe, especially at higher frequencies for which the guided mode becomes more tightly confined in the waveguide central layers such that a slight position offset of the probe with respect to the actual peak field amplitude can significantly lower the detected signal. A similar remark can be made regarding the exact location of the incident beam focus spot on the waveguide input cross section.

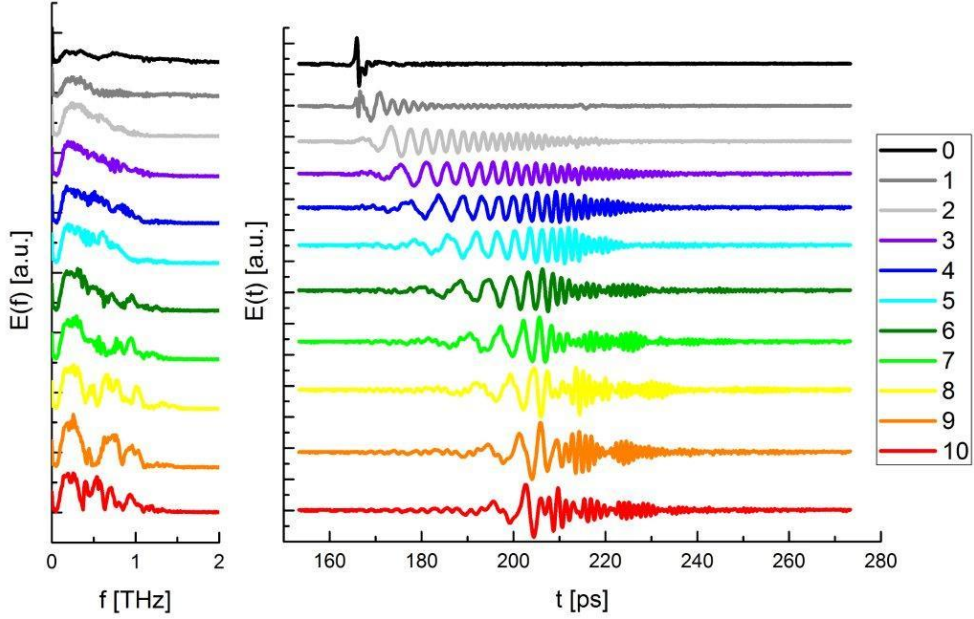

Fig. 5. (a) Experimentally measured electric field in frequency domain at the output of the waveguide for different number of layers constituting the waveguide. Black line – 0 layers, dark gray – 1 layer, light gray – 2 layers, purple – 3 layers, blue – 4 layers, cyan – 5 layers, dark green - 6 layers, light green – 7 layers, yellow – 8 layers, orange – 9 layers, red – 10 layers. (b) Experimentally measured electric field in time domain at the output of the waveguide for different number of layers constituting the waveguide

We can now derive an expression for the intensity of the transmitted field as measured by the near field detector:

$$T(\omega) = \left|\vec{E}_{output}(x_0, y_0, \omega)\right| = \left|\sum_{m=1}^{N} C_m \cdot \vec{E}_m(x_0, y_0, \omega) e^{i\frac{\omega}{c} n_{eff,m} \cdot L_w} \cdot e^{-\frac{\alpha_m L_m}{2}}\right| \quad (1)$$

where $(x_0, y_0)$ denotes the coordinates of the waveguide cross-section center.

In Fig. 5 we present experimentally measured electric field at the output of the waveguide for different number of layers constituting the waveguide. We note that increasing of the number of layers monotonically increases the effective guiding range of the waveguide, from 0.15 – 0.6 THz for one layer waveguide to 0.15 – 2.0 THz for ten layers waveguide. Not regarding the number of layers all the spectra exhibit small oscillations of transmission. And also there some big dips in transmission spectra, their amounts depend on the number of layers constituting the waveguide. There is just one transmission peak for waveguides with 1 to 3 layers, two transmission peaks for 4 to 6 layers and three peaks for the amount of layers equal to 7 and higher.

To study the propagation in the waveguide numerically we have imported its cross-section geometries into COMSOL Multiphysics FEM software, and then solved for the complex effective refractive indices and field profiles. Coupling coefficients have been computed for each mode as overlap integrals of the modal field distribution and the initial Gaussian along vertical axis and uniform along horizontal axis field distribution. Finally, transmission through the waveguide has been calculated using Equation (1) and taking into account reference value of the EM field.

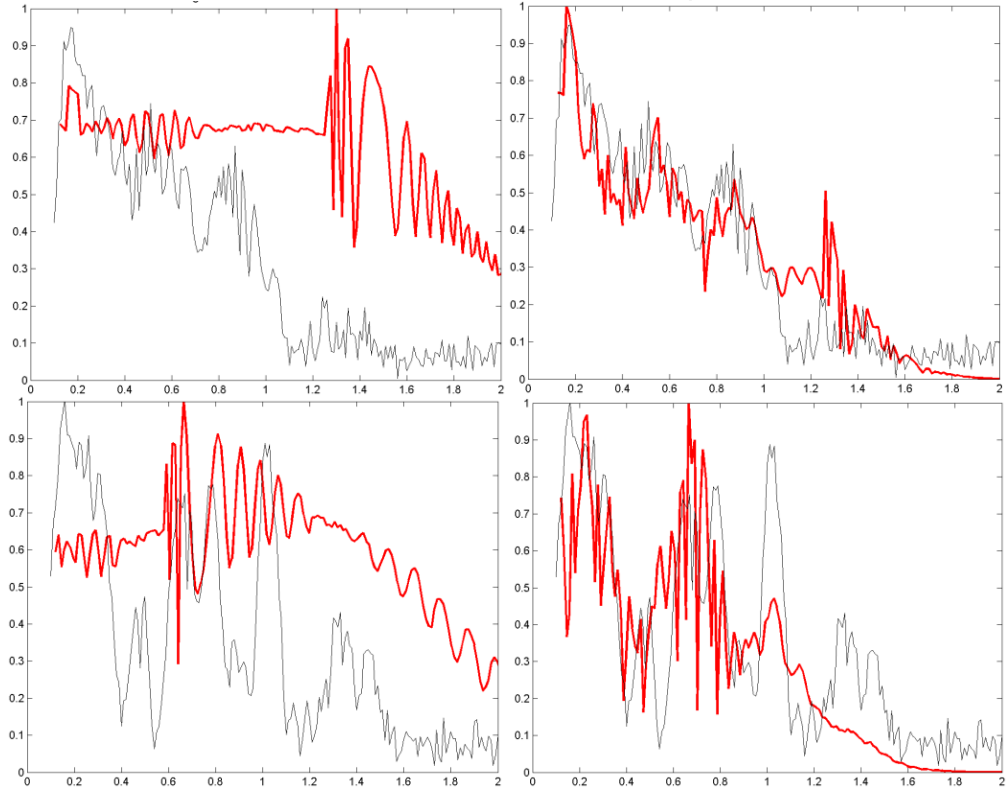

Fig. 6. Black line – normalized experimental transmittance, red line – theoretical transmittance (a) Simulation of 4 layers after 11 cm of propagation (b) Simulation of 4 layers after 35 cm of propagation, (c) Simulation of 8 layers after 11 cm of propagation (d) Simulation of 8 layers after 35 cm of propagation

In horizontal plane the waveguide is limited by the metal walls, making possible the reflection of the transmitted light from them and leading to the appearance of the numerous so-called side modes. For each mode which is characterized by the number of its intensity maximums in the vertical direction of the waveguide there is a series of corresponding side modes with slightly different horizontal angles of propagation, which in their turn are characterized by the number of intensity peaks in horizontal direction of the waveguide. It has to be mentioned that not all of these modes can be excited using our configuration of initial beam field but only odd ones having coupling coefficients different from zero. This variety of side modes leads to the appearance of small oscillation on the transmittance curves mentioned above.

And according to the numerical simulation the appearance of the big dips on the transmittance graphs corresponds to cut-off frequencies of higher-order modes of the waveguides, this time not side modes but modes with different number of peaks in vertical direction. Again, only modes with odd number of peaks can be excited with Gaussian/uniform initial beam configuration. Coupling coefficient into higher order modes is relatively high for this waveguide, thus, a significant fraction of power goes into higher-order modes leading to inter-mode interference effect at the output of the waveguide.

The results of the numerical simulation of the transmission through the waveguide are depicted in Fig. 6 for several values of waveguide layers. The presence of the transmittance oscillations and the dips due to appearance of the higher-order modes are well described by the model. However, for good agreement between the experimental and theoretical values of transmittance the length of the waveguide in simulation has to be increased up to 35 cm compared to initial length of 11 cm. The absorption losses of the waveguide turn out to be significantly higher than the results of the simulation. Comparison between Fig. 6 (a) and (b) for 4 layers and Fig.6 (c) and (d) for 8 layers illustrates this problem.

**Conclusion**

Planar dielectric subwavelength waveguides are proposed as waveguides for THz radiation. The waveguide design maximizes the fraction of power guided in the air to avoid the complexity that all the materials are highly absorbent in the THz region. Convenient access to the mode region makes them also useful for sensing of biological and chemical specimens in the THz region. The transmission and absorption properties of such waveguides have been investigated both experimentally using THz-TDS spectroscopy and theoretically using finite element software. Good agreement between experimental data and theoretical results has been achieved.

**Acknowledgement**

M. Skorobogatiy thanks Prof. H. Ito and Dr. H. Minamide for hosting him in Sendai during his sabbatical in 2010.